\documentstyle[aps,prl,amsfonts]{revtex}
\begin{document}

\newcommand{\N}{{\Bbb N}}
\newcommand{\Z}{{\Bbb Z}}
\newcommand{\R}{{\Bbb R}}
\newcommand{\w}{{\omega}}
\newcommand{\QP}{{\em QP }}
\newcommand{\ZQP}{{\em ZQP }}
\newcommand{\Tr}{{\mathrm{Tr}}}
\newcommand{\QFT}{{\cal F}}
\newcommand{\cT}{{\cal T}}
\newcommand{\cU}{{\cal U}}
\newcommand{\cJ}{{\cal J}}
\newcommand{\cS}{{\cal S}}
\newcommand{\cR}{{\cal R}}
\newcommand{\cI}{{\cal I}}
\newcommand{\cW}{{\cal W}}
\newcommand{\cV}{{\cal V}}
\newcommand{\cH}{{\cal H}}

\draft
\title{Function-dependent Phase Transform in Quantum Computing}
\author{Dong Pyo Chi,\cite{dpchi}
        Jinsoo Kim,\cite{jskim} and
        Soojoon Lee\cite{level}
}
\address{Department of Mathematics, Seoul National University,
 Seoul 151-742, Korea}
\date{\today}
\pagestyle{plain}
\pagenumbering{arabic}
\maketitle

\begin{abstract}
We construct a quantum algorithm
that performs function-dependent phase transform and
requires no initialization of an ancillary register.
The algorithm recovers the initial state of an ancillary register
regardless of whether its state is pure or mixed.
Thus we can use any qubits as an ancillary register
even though they are entangled with others and
are occupied by other computational process.
We also show that our algorithm is optimal in the sense of
the number of function evaluations.
\end{abstract}



Quantum computation is based on three quantum phenomena:
superposition of states, quantum interference, and quantum entanglement.
These effects enable exponential speedups in the solutions of certain
problems and allow one to transgress some boundaries of classical
computational complexity theory
\cite{Deutsch,DJ,BB92,BV,Simon,Shor,BB94,Grover,BBHT,BBBV,BH}.
Most known quantum algorithms \cite{Deutsch,DJ,BV,Grover,BH,CK}
rely on conditional phase transform
the realization of which is accomplished by
the quantum Fourier transform or the Walsh-Hadamard operator
together with the unitary operator evaluating a given function
or quantum oracle.
In general, conditional phase transform
can be described by the operation
$\sum_{x=0}^{N-1} \alpha_x \left|{x}\right\rangle \mapsto
\sum_{x=0}^{N-1} \exp[2\pi i f(x)/M] \alpha_x \left|{x}\right\rangle$
for a function $f:\Z_N \rightarrow \Z_M$,
which we call {\em $f$-dependent phase transform}.
The resulting interference pattern
facilitates determining global property of the underlying function.

We need a quantum circuit evaluating a function
to perform function-dependent phase transform and
unitary evolution of quantum computational process requires an ancillary
register from which we have to extract the desired relative phases
conditioned on the given function.
All previous quantum algorithms resort to
initialization of the ancillary register before the computation.
We may ask a question:
Is it possible to perform function-dependent phase transform
without initializing and deforming the state of the ancillary register?
If it were possible,
energy dissipation caused by initialization process would be avoidable
and any register that could contain useful information to be preserved
could temporarily be used as an ancillary register.
In this work we construct a quantum algorithm that implements
function-dependent phase transform
without initializing an ancillary register.
Furthermore, the application of the constructed algorithm
retrieves the initial state of the ancillary register.
Thus the ancillary register can consist of
any qubits collected from any other registers even though
they are being used in other computation
which can proceed after carrying out
their auxiliary duty in function-dependent phase transform.
We show that to realize function-dependent phase transform
at least two operations dependent on the given function are necessary.
Thus the presented algorithm is optimal
in the sense that it involves only two function evaluations.
Of course, if any kind of initialization is involved,
one function evaluation is sufficient.


The $f$-dependent phase transform
$\cR_{k,f}:\left|{x}\right\rangle \mapsto
\w_M^{kf(x)} \left|{x}\right\rangle $ plays an important
role in quantum algorithms
where
$\w_M=\exp(2\pi i /M)$ is a primitive $M$-th root of unity and
$k \in \Z_M$ may be chosen appropriately depending on the given problems.
For simplicity, we assume that $N$ and $M$ are powers of 2,
that is, $N=2^n$ and $M=2^m$ for some nonnegative integers $n$ and $m$.
In order for the information on the given function
to be encoded in the phases
it is necessary to evaluate the given function on quantum computer.
On quantum computer the evaluation of a function is performed by
the unitary operation
$\cU_f: \left|{x}\right\rangle \otimes \left|{y}\right\rangle \mapsto
\left|{x}\right\rangle \otimes \left|{y+f(x)}\right\rangle$
for $x \in \Z_N$ and $y \in \Z_M$.
The first $n$-qubit register we call the {\em control register}
contains the states we wish to interfere.
The second $m$-qubit register called the {\em function} or
{\em ancillary register}
is used to draw relative phase changes in the first register.
The superposition principle of quantum mechanics allows
us to prepare the computer in a coherent superposition of input states
and to compute exponentially many values of $f$ in superposition
with a single application of $\cU_f$.
This phenomenon is the basis for quantum parallelism
which leads to a completely new model of computation.
In view of the second register the function evaluation adopts
a translation operator
$\cT_z : \left|{y}\right\rangle \mapsto
\left|{y+z}\right\rangle$ where
$z$ is dependent on the state of the first register.
The operator $\cU_f$ can be described in terms of the translation
operator on the second register;
\begin{equation}\label{eq:Uf}
\cU_f: \sum_{x,y} \alpha_{xy}
\left|{x}\right\rangle \otimes \left|{y}\right\rangle \mapsto
\sum_{x,y} \alpha_{xy} \left|{x}\right\rangle \otimes \cT_{f(x)} \left|{y}\right\rangle  .
\end{equation}
We note that to implement $\cR_{k,f}$ at least two registers
are necessary due to $\cU_f$.
By adding an ancillary register
the effect of the $\cR_{k,f}$ on the control register
can be viewed as phase changes in the ancillary register
dependent on the states of the control register.
To be more specific, we define a unitary operator
$\cJ_{k,z}: \left|{y}\right\rangle \mapsto
\w_M^{kz} \left|{y}\right\rangle $ for $y\in \Z_M$.
Then $\cR_{k,f} \otimes I$ can explicitly be written by
\begin{equation}\label{eq:R}
 \cR_{k,f} \otimes \cI :
 \sum_{x,y} \alpha_{xy} \left|{x}\right\rangle \otimes
 \left|{y}\right\rangle \mapsto
 \sum_{x,y} \alpha_{xy} \left|{x}\right\rangle \otimes
 \cJ_{k,f(x)} \left|{y}\right\rangle .
\end{equation}
We note that
$\cJ_{k,z}$ has one eigenvalue $\w_M^{kz}$ and the corresponding
eigenspace is the whole Hilbert space.
Due to the expressions (\ref{eq:Uf}) and (\ref{eq:R})
we can concentrate on the operations of the ancillary register.

Especially when $f$ is the identity map $\cI$,
$\cR_{k,\cI}$ maps $\left|{y}\right\rangle$ to
$\w_M^{ky} \left|{y}\right\rangle$,
in which the phase-encoded information depends on its state,
and can be obtained by $\QFT^\dagger \cT_k^\dagger \QFT$
where $\QFT$ is the quantum Fourier transform.
A quantum algorithm to implement $\cJ_{k,z}$ can be realized
using $\cR_{k,\cI}$ and $\cT_z$.
We prepare an arbitrary $m$-qubit register
whose state is
$\left|{\Psi}\right\rangle
= \sum_{y=0}^{M-1} \alpha_y\left|{y}\right\rangle$
and proceed the following algorithm:
(i) Apply $\cT_z$.
(ii) Apply $\cR_{k,\cI}$.
(iii) Apply $\cT_z^\dagger=\cT_{-z}$.
(iv) Apply $\cR_{k,\cI}^\dagger=\QFT^\dagger \cT_{k} \QFT$.
Then the state evolves as follows:
\begin{eqnarray}
\left|{\Psi}\right\rangle
&\stackrel{\cT_z}{\longrightarrow}&
 \sum_{y=0}^{M-1}\alpha_y \left|{y+z}\right\rangle \nonumber \\
&\stackrel{\cR_{k,\cI}}{\longrightarrow}&
 \sum_{y=0}^{M-1} \w_M^{k(y+z)} \alpha_y \left|{y+z}\right\rangle \nonumber \\
&\stackrel{\cT_z^\dagger}{\longrightarrow}&
 \sum_{y=0}^{M-1} \w_M^{k(y+z)} \alpha_y \left|{y}\right\rangle  \nonumber \\
&\stackrel{\cR_{k,\cI}^\dagger}{\longrightarrow}&
 \w_M^{k z}\left|{\Psi}\right\rangle . \label{Alg:J}
\end{eqnarray}
Therefore we get
$\cR_{k,\cI}^\dagger \cT_z^\dagger \cR_{k,\cI} \cT_z
\left|{\Psi}\right\rangle
= \w_M^{kz} \left|{\Psi}\right\rangle $
for an arbitrary $\left|{\Psi}\right\rangle $, namely, we have
\begin{equation}\label{eq:commutator}
\cJ_{k,z} = \w_M^{k z} \cI
= \cR_{k,\cI}^\dagger \cT_z^\dagger \cR_{k,\cI} \cT_z .
\end{equation}

{\em Theorem 1.} There exists a quantum algorithm to implement $\cJ_{k,z}$
using two $\cT_{\pm z}$.

The algorithm for $\cJ_{k,z}$ is not unique.
All cyclic permutations of the steps are identical.
For example, we can start at Step (ii), perform successive steps, and
end at Step (i).
In fact, if we use the notation $[A,B]=ABA^{-1}B^{-1}$
then by Eq.\ (\ref{eq:commutator}) we have
\begin{eqnarray}
\cJ_{k,z}
 &=& [\cR_{k,\cI}^\dagger,\cT_z^\dagger]
  = [\cT_z,\cR_{k,\cI}^\dagger] \nonumber \\
 &=& [\cR_{k,\cI},\cT_z]
  = [\cT_z^\dagger,\cR_{k,\cI}] \label{eq:commutator2}
\end{eqnarray}
with its inverse
$\cJ_{-k,z} = [\cR_{k,\cI},\cT_z^\dagger] = [\cT_z,\cR_{k,I}]
 = [\cR_{k,\cI}^\dagger,\cT_z] = [\cT_z^\dagger,\cR_{k,\cI}^\dagger]$.
Furthermore,
noting that $\cS_{k,\cI} = \QFT \cT_k^\dagger \QFT$ maps
$\left|{y}\right\rangle $ to $\w_M^{ky} \left|{-y}\right\rangle $
one can easily check that
\begin{equation}\label{eq:J}
\cJ_{k,z} = \cS_{k,\cI} \cT_z \cS_{k,\cI} \cT_z ,
\end{equation}
which also offers another implementation.
We remark that $\cS_{k,\cI}^\dagger = \cS_{k,\cI}$.
Therefore there are many methods to implement $\cJ_{k,z}$.
However, the number of $\cT_{\pm z}$ in each implementation is
always equal to two and cannot be reduced.

Let us suppose that there exists a quantum algorithm
implementing $\cJ_{k,z}$
where the only way to implement the given information on $z$
is through $\cT_{\pm z}$.
Then the dependence of $z$ requires at least one $\cT_{\pm z}$
at a certain step and hence
the overall unitary operation performed by the algorithm can be
written by $\cV_2 \cT_{\pm z} \cV_1 = \w_M^{kz} \cI$
for some unitary operators $\cV_1$ and $\cV_2$.
Since $\cV_1 \cV_2 \cT_{\pm z} = \w_M^{kz} \cI$, it
is enough to consider a unitary operator $\cV$ such that
$\cV \cT_{\pm z} = \w_M^{kz} \cI$.
Since $\cV = \w_M^{kz} \cT_{\pm z}^\dagger$, $\cV$ depends on $z$.
Thus in some another step of the algorithm we have to use information
on $z$ once more and so the overall procedure includes
at least two translations by $\pm z$.
This observation will later be used in showing that
our algorithm for $\cR_{k,f}$ is optimal.


We now turn to the $f$-dependent phase transform $\cR_{k,f}$.
We let
$\left|{\Phi}\right\rangle
= \sum_{x=0}^{N-1} \alpha_x \left|{x}\right\rangle$ and
$\left|{\Psi}\right\rangle
= \sum_{y=0}^{M-1} \beta_y\left|{y}\right\rangle$
be the respective states of the control and the ancillary registers.
It is noted that no initialization is involved
during the preparation of the registers.
By inspecting Eqs.\ (\ref{eq:Uf}) and (\ref{eq:commutator})
the algorithm (\ref{Alg:J}) for $\cJ_{k,z}$
leads to an algorithm for $\cR_{k,f}$:
(i) Apply $\cU_f$.
(ii) Apply $\cI \otimes \cR_{k,\cI}$.
(iii) Apply $\cU_f^\dagger = \cU_{-f}$.
(iv) Apply $\cI \otimes \cR_{k,\cI}^\dagger = \cI \otimes \cR_{-k,\cI}$.
This procedure makes the state of the registers evolve as follows:
\begin{eqnarray}
\left|{\Phi}\right\rangle \otimes \left|{\Psi}\right\rangle
 &\stackrel{\cU_f}{\longrightarrow}&
 \sum_{x=0}^{N-1} \sum_{y=0}^{M-1}
  \alpha_x \beta_y \left|{x}\right\rangle \otimes
  \left|{y+f(x)}\right\rangle  \nonumber \\
 &\stackrel{\cI \otimes \cR_{k,\cI}}{\longrightarrow}&
 \sum_{x=0}^{N-1} \sum_{y=0}^{M-1}
  \alpha_x \beta_y \w_M^{k(y+f(x))} \left|{x}\right\rangle \otimes
  \left|{y+f(x)}\right\rangle  \nonumber \\
 &\stackrel{\cU_f^\dagger}{\longrightarrow}&
 \sum_{x=0}^{N-1} \sum_{y=0}^{M-1}
  \alpha_x \beta_y \w_M^{k(y+f(x))} \left|{x}\right\rangle \otimes
  \left|{y}\right\rangle  \nonumber \\
 &\stackrel{\cI \otimes \cR_{k,\cI}^\dagger}{\longrightarrow}&
 \left( \sum_{x=0}^{N-1} \w_M^{k f(x)}
  \alpha_x \left|{x}\right\rangle  \right)
  \otimes \left|{\Psi}\right\rangle  . \label{Alg:R}
\end{eqnarray}
Now we discard the ancillary register.
Then we obtain the $f$-dependent phase transform
$\cR_{k,f} : \left|{x}\right\rangle
\mapsto \w_M^{kf(x)} \left|{x}\right\rangle$.

{\em Theorem 2.} There exists a quantum algorithm
that implements function-dependent phase transform
using two evaluations of a given function
such that the ancillary register preserves its initial state.

Since the $f$-dependent phase transform $\cR_{k,f}$ can be written
in terms of $\cJ_{k,z}$ as in Eq.\ (\ref{eq:R})
and by Eq.\ (\ref{eq:commutator2})
there are many methods to realize $\cJ_{k,z}$,
we can conclude that the algorithm for $\cR_{k,f}$ is also not unique.

In the procedure (\ref{Alg:R}) we have assumed that
an ancillary register is in a pure state.
However, this is not an essential requirement.
In fact, any mixed state is allowed.
To be more precise,
let $A$ be a quantum system to be used as an ancillary register
and its state be described by the density operator $\rho^A$.
Then there exists a {\em reference system} $R$ such that
the compound system $AR$ is in pure entangled state
$\left|{\Psi^{AR}}\right\rangle $
that gives rise to the given reduced state $\rho^A = \Tr_R(\rho^{AR})$
where
$\rho^{AR} = \left|{\Psi^{AR}}\right\rangle
 \left\langle \Psi^{AR} \right|$
is called {\em purification} of $\rho^A$.
Using the Schmidt decomposition we can rewrite
$\left|{\Psi^{AR}}\right\rangle $ as
$\sum_{y=0}^{M-1} \alpha_y \left|{y^A}\right\rangle \otimes
\left|{\Psi_y^R}\right\rangle $.
We note that the states $\left|{\Psi_y^R}\right\rangle$'s
may not form the standard basis
for the subsystem $R$ but just an orthonormal basis
while the states $\left|{y^A}\right\rangle$'s form the standard basis
for the subsystem $A$.
Now applying the above algorithm to
$\left|{\Phi}\right\rangle \otimes \left|{\Psi^{AR}}\right\rangle$
one can see that the final state becomes
$\left( \cR_{k,f} \left|{\Phi}\right\rangle  \right) \otimes
\left|{\Psi^{AR}}\right\rangle$.
Thus our algorithm works whether the state of the ancillary register
is pure or mixed.
This implies that
we can compose an ancillary register of
any $m$ qubits which are collected out of any other registers
even though they are still being used in other computational process
and are possibly entangled with other qubits.
The presented algorithm (\ref{Alg:R}) recovers
the initial state of the joint system $AR$
after extracting the desired relative phase changes.
Thus the qubits in the temporarily composed register can be
restored to their positions to continue the suspended computation.

Our algorithm requires two evaluations of $f$,
i.e., $\cU_f$ and $\cU_f^\dagger$
[or two $\cU_f$ when Eq.\ (\ref{eq:J}) is applied].
This is because we employ no initialization.
We know that $\cU_f$ causes translations in the ancillary register
by Eq.\ (\ref{eq:Uf}) and
that any quantum algorithm for $\cJ_{k,z}$
adopts at least two translations.
Thus we have the following theorem.

{\em Theorem 3.} Any quantum algorithm that implements
function-dependent phase transform
without initialization
requires at least two evaluations of a given function.

On the other hand, if the ancillary register is initializable
only one evaluation of $f$ is sufficient.
Indeed it is clear that $\QFT \left|{-k}\right\rangle $
is an eigenvector of $\cT_z$
with the corresponding eigenvalue $\w_M^{kz}$.
If we let
$\left|{\Psi}\right\rangle  = \QFT \cT_{-k} \left|{0}\right\rangle$,
then $\cU_f$ maps
$\left|{x}\right\rangle \otimes \left|{\Psi}\right\rangle$ to
$\w_M^{k f(x)} \left|{x}\right\rangle \otimes \left|{\Psi}\right\rangle$.
The special case for $k=1$ was studied in \cite{Cleve1,Cleve2}.

Initialization in general sense is a process to transform
the state of a quantum system to a definite pure state,
which can later be rotated to $\left|{0}\right\rangle $ as usual
or any other desired state by frame change.
When we are to initialize the subsystem $A$
we cannot avoid corrupting the correlation between
the subsystems $A$ and $R$,
which can be measured by the quantum mutual entropy $S(A:R)$.
If the subsystem $A$ is entangled with the reference system $R$,
that is, $S(A:R)\ne 0$,
then even when $\rho^A$ is known
we cannot initialize the subsystem $A$
by local unitary operations on $A$.
We note that the quantum mutual entropy is invariant under
local unitary operations of product form for each subsystem.
If the bipartite systems $A$ and $R$ are separable,
that is, $S(A:R) = 0$,
then a certain frame change on the subsystem $A$
effects on initialization of the ancillary register
without knowing the total state $\rho^{AR}$.
In this sense we say that the subsystem $A$ is
{\em nondestructively initializable}
when $\rho^A$ is pure and known
and initialization by a local frame change on the subsystem $A$
is called {\em nondestructive initialization}.
If nondestructive initialization is adopted
then the ancillary register regains its early state
and one evaluation of a function is sufficient for
function-dependent phase transform.

Let us consider a more general function
$f:\Z_N \rightarrow [0,1) \subset \R$.
We define $m$-bit approximation $\tilde{f} : \Z_N \rightarrow \Z_M$
of $f$ by $\tilde{f}(x)=\sum_{i=1}^{m}a_i 2^{m-i}\in \Z_M$
for $x \in \Z_N$
where $(0. a_1 a_2 \dots a_m)_2= \sum_{i=1}^{m}a_i 2^{-i}$
is an $m$-bit binary expansion of $f(x)$
for $a_i \in \Z_2$.
Then the {\em approximate $f$-dependent phase transform}
$\cR_{k,\tilde{f}}$ approximates the operation
$\left|{x}\right\rangle \mapsto
e^{2\pi i f(x)} \left|{x}\right\rangle $ \cite{Cleve1,Cleve2}.
This approximate $f$-dependent phase transform
can be applied to the conditional $\gamma$-phase transform and
the $\beta$-phase diffusion transform in \cite{CK}.


All known quantum algorithms are based on the effect of
function-dependent phase transform and
the quantum Fourier transform (or the Walsh-Hadamard operator).
The quantum Fourier transform enables
one to find the period of a function in polynomial time
and plays an essential role in Shor's
quantum polynomial-time algorithms \cite{Shor}
for the integer factoring and
the discrete logarithm problems which are known to be intractable
on classical computer.
It also enables us to construct function-dependent phase transform
without initializing an ancillary register,
which can immediately be applied to most known quantum algorithms.

Deutsch and Jozsa \cite{Deutsch,DJ} presented a simple promise problem
to determine whether a Boolean function
$f:\Z_N \rightarrow \Z_2$ is either constant or balanced
and showed that it can be solved efficiently without error
on quantum computer
while it requires exhaustive search
to solve deterministically without error in a classical setting.
The key of their algorithm is
the $\pi$-rotation of phases controlled by the query result of
quantum oracle.
In this problem $M=2$ and $k=1$.
Then $\QFT$ becomes the Walsh-Hadamard operator
$\cW=\scriptsize \frac{1}{\sqrt{2}}
\pmatrix{1 &\hphantom{-} 1\cr 1 & -1}$
and $\cT_1=\cT_{-1}$ becomes the Pauli spin operator
$\sigma_x = \scriptsize\pmatrix{ 0 & 1 \cr 1 & 0 }$
which represents a bit-flip.
The operator $\cR_{1,\cI}$ is a phase-flip operator
$\sigma_z = \scriptsize\pmatrix{ 1 & \hphantom{-} 0 \cr 0 & -1}$.
Thus the overall scheme for $\cR_{1,f}$ is
$(\cI\otimes\sigma_z) \cU_f (\cI\otimes\sigma_z) \cU_f$ \cite{Kim}.
Grover constructed a quantum algorithm that can find a particular item
in expected time $O(\sqrt{N})$ when an unstructured list of $N$
items are given \cite{Grover,BBHT,BBBV}.
His algorithm relies on the conditional phase transform
$S_f: \left|{x}\right\rangle  \mapsto (-1)^{f(x)} \left|{x}\right\rangle$
and the diffusion transform $D = \cW S_0 \cW^\dagger$
where $f$ is the Boolean function computed by an oracle,
$S_0 = S_{f_0}$, and $f_0(x) = \delta_{0x}$.
In this case $S_f=\cR_{1,f}$ with $M=2$.
Brassard and H{\o}yer \cite{BH} combined
Simon's \ZQP algorithm \cite{Simon} and Grover's quantum search algorithm
and showed that
Simon's problem can be solved on a quantum computer
in worst-case polynomial time and thus is in \QP class.
Thus their \QP algorithm mainly depends on the conditional
phase transform.
Chi and Kim \cite{CK} generalized Grover's algorithm
and showed that a quantum computer can search a database
by a single query when the number of solutions is equal to or
more than a quarter.
Their algorithm makes use of
the conditional $\gamma$-phase transform
$S_{f,\gamma} :\left|{x}\right\rangle \mapsto
e^{i\gamma f(x)} \left|{x}\right\rangle$
and the $\beta$-phase diffusion transform
$D_{\beta} = W_l S_{l,\beta} W_l^\dagger$
where $W_l$ is any unitary transformation satisfying
$W_l \left|{l}\right\rangle
= {\scriptsize\frac{1}{\sqrt{N}}}
\sum_{x=0}^{N-1} \left|{x}\right\rangle$
and $S_{l,\beta} = S_{f_l,\beta}$ with $f_{l}(x)= \delta_{lx}$.
In this case we can use approximate function-dependent phase transform.

All conditional phase transforms fall into
the category of function-dependent phase transform.
Therefore our algorithm for function-dependent phase transform
is directly applicable to most quantum algorithms.
We note that for general positive integers $N$ and $M$
the approximate Fourier transform in \cite{Kitaev} can be used
in our algorithm.

In summary, we generalized conditional phase transform
to function-dependent phase transform
and presented a quantum algorithm that performs function-dependent
phase transform and does not require any kind of initialization of
an ancillary register.
Our algorithm recovers the initial state of the ancillary register.
Thus we can compose an ancillary register of any qubits
regardless of whether they are entangled with others or
being used in another computational process.
Our algorithm employs two evaluations of a given function
and is optimal in that
any quantum algorithm that implements function-dependent phase
transform without initialization
requires at least two evaluations of a given function.

This work was supported by the Brain Korea 21 Project.


\end{document}